\documentclass[twocolumn,dvipsnames]{aastex62}

\usepackage{CJK}
\usepackage{framed,amsmath,bm}

\newcommand\vtheta{\bm{\theta}}

\newcommand\betay{y_0}
\newcommand\fyi{f_y^i}
\newcommand\fyij{f_y^{ij}}

\newcommand\ebmy{y(\vtheta)}
\newcommand\np{n_\mathrm{p}}

\def\fnu{\mathcal{F}_\nu}
\def\gnu{\mathcal{G}_\nu}

\def\dim#1{{\rm\,#1}}
\def\Lya{Ly$\alpha$}
\def\HI{{\rm H\scriptstyle\rm I}}
\def\HII{{\rm H\scriptstyle\rm II}}

\graphicspath{{./}{figures/}}

\received{January 8, 2024}
\revised{August 22, 2024}
\accepted{September 8, 2024}
\submitjournal{ApJ}

\shorttitle{}
\shortauthors{Zhu, Gnedin, Avestruz}

\begin{document}
\begin{CJK*}{UTF8}{gkai}

\title{On the Physical Nature of \Lya\ Transmission Spikes in High Redshift Quasar Spectra}

\correspondingauthor{Hanjue Zhu (朱涵珏)}
\email{hanjuezhu@uchicago.edu}
\author[0000-0003-0861-0922]{Hanjue Zhu (朱涵珏)}
\affiliation{Department of Astronomy \& Astrophysics; 
The University of Chicago; 
Chicago, IL 60637, USA}

\author[0000-0001-5925-4580]{Nickolay Y.\ Gnedin}
\affiliation{Theory Division; 
Fermi National Accelerator Laboratory;
Batavia, IL 60510, USA}
\affiliation{Kavli Institute for Cosmological Physics;
The University of Chicago;
Chicago, IL 60637, USA}
\affiliation{Department of Astronomy \& Astrophysics; 
The University of Chicago; 
Chicago, IL 60637, USA}

\author[0000-0001-8868-0810]{Camille Avestruz}
\affiliation{Department of Physics; University of Michigan, Ann Arbor, MI 48109, USA}
\affiliation{Leinweber Center for Theoretical Physics; University of Michigan, Ann Arbor, MI 48109, USA}

\begin{abstract}
We investigate Lyman-alpha (\Lya) transmission spikes at $5.2 < z < 6.8$ using synthetic quasar spectra from the ``Cosmic Reionization on Computers" simulations. We focus on understanding the relationship between these spikes and the properties of the intergalactic medium (IGM). Disentangling the complex interplay between IGM physics and the influence of galaxies on the generation of these spikes presents a significant challenge. To address this, we employ Explainable Boosting Machines, an interpretable machine learning algorithm, to quantify the relative impact of various IGM properties on the \Lya\ flux. Our findings reveal that gas density is the primary factor influencing absorption strength, followed by the intensity of background radiation and the temperature of the IGM. Ionizing radiation from local sources (i.e.\ galaxies) appears to have a minimal effect on \Lya\ flux. The simulations show that transmission spikes predominantly occur in regions of low gas density. Our results challenge recent observational studies suggesting the origin of these spikes in regions with enhanced radiation. We demonstrate that \Lya\ transmission spikes are largely a product of the large-scale structure, of which galaxies are biased tracers. \\
\end{abstract}

\section{Introduction}\label{sec:intro}

Observations of the Lyman-alpha (\Lya) forest in quasar spectra effectively map hydrogen gas distribution in the universe. Simulations from the mid-1990s demonstrated that the \Lya\ forest at intermediate redshifts ($z\sim2-4$) manifests the small-scale tail of the cosmic large-scale structure, challenging earlier physical models that attributed the features to astronomical objects such as pressure-confined clouds, shocks, and mini-halos \citep{Lya1,Lya2,Lya3}. The \Lya\ forest has subsequently become a fundamental tool in modern astrophysics, with applications ranging from measuring cosmological parameters and matter clustering to constraining the level of turbulence in the intergalactic medium (IGM) \citep{Gaikwad2017,Bolton2022}, as well as the thermal and ionization state of the gas \citep{Hui1997,Becker2011,Becker2013,Boera2016,Telikova2019,Walther2019,Gaikwad2020,Gaikwad2021}.

However, the large \Lya\ cross-section means that even a tiny hydrogen neutral fraction ($x_\HI \sim 10^{-4}$) produces complete absorption \citep{Bi1997,Rauch1998,fgpa2}. As the universe is more neutral at higher redshifts, the Lyman-alpha forest becomes denser. Eventually, somewhere around $z\sim5$, the appearance of the quasar spectrum changes dramatically - clear absorption lines disappear, replaced by blended absorption features and sporadic ``transmission spikes'', i.e.\ regions of incomplete absorption \citep[see][for a review]{Becker2015}. The quasar absorption spectrum now resembles an emission spectrum, yet the transmission spikes do not exhibit the defined shapes typical of Doppler or Voigt profiles - in fact, the precise shapes and variations of the transmission spikes are still unknown. The origins of these spikes are also unclear: while many of the spikes likely arise from low-density regions (where there is minimum absorption)\citep{fgpa1,Miralda-Escude2000,Peeples2010}, it is uncertain if they are exclusively associated with cosmic voids \citep{Garaldi2019,Keating2020,Nasir2020,Gaikwad2020}. Moreover, the spatial correlation between the transmission spikes and galaxies is unclear. While galaxies enhance local ionizing radiation, leading to decreased absorption \citep{Adelberger2003,Cantalupo2012}, they also live in denser regions where absorption is naturally higher \citep{Mo1996,Rakic2012}. Despite this, contradictory evidence exists in current observational data, which shows decreased absorption near galaxies.\citep[e.g.][]{Kakiichi2018,Meyer2019}. However, it is important to note that these observations are sensitivity-limited and potentially missing low surface brightness galaxies that could significantly contribute to the ionizing photon budget.

Our understanding of the \Lya\ forest is rather comprehensive at $z<5$, yet it becomes markedly limited beyond $z\sim5$. The influx of high-quality data from JWST intensifies this discrepancy. In the near future, 30-meter class telescopes will begin observing fainter quasars in fields with sufficiently deep JWST exposures, and it is anticipated that numerous synergistic observational programs will emerge \citep{LaPlante2019, Furlanetto2019, Rieke2019, Cooray2019, Becker2019}. Such programs could provide valuable observational constraints and enhance our understanding of the connection between galaxies and the IGM. However, these detailed observational data risk being under-leveraged without a proper theoretical interpretation.

Fortunately, reliable theoretical and computational tools are now available to enhance our understanding of high-redshift quasar spectra significantly. State-of-the-art cosmological simulations of cosmic reionization produce reasonably realistic models in volumes in excess of 100 cMpc, providing the foundation for creating high-resolution synthetic quasar spectra \citep{gnedin14,coda1,coda2,thesan,thesan2}. The ``Cosmic Reionization on Computers'' (CROC) project \citep{gnedin14,gnedinandkaurov_14} is one of them. In this work, we rely on CROC simulations to provide a reasonably realistic model for the high redshift IGM. Furthermore, this study incorporates the use of Explainable Boosting Machines (EBMs), an enhanced iteration of Generalized Additive Models (GAMs) as implemented by \citealt{ebm}.  GAMs are built on the principle that the model's output is the aggregate of individual contributions from each feature, offering interpretability by revealing feature significance in predicting outcomes. EBMs enhance this interpretability by integrating modern machine learning techniques to boost performance without compromising the ability to understand the models' decisions. The synergy of advanced reionization simulations with cutting-edge machine learning tools equips us to identify the properties of the environments from which transmission spikes in $z>5$ quasar spectra originate.

This paper is organized as follows. Section~\ref{sec:methods} describes the CROC simulations, synthetic \Lya\ spectra generation, and EBM methodology. In Section~\ref{sec:results}, we examine model performances with different sets of IGM properties as inputs, as well as the average contribution of each parameter to the target quantity (\Lya\ transmission flux). We summarize and discuss these results in Section~\ref{sec:discussion}.

\section{Methodology}\label{sec:methods}

\subsection{CROC Simulations}

In this study, we employ simulations from the Cosmic Reionization on Computers (CROC) project, a suite of cosmological hydrodynamic simulations of cosmic reionization. For detailed information on the simulations, we direct readers to the CROC methods paper \citep{gnedin14}. We utilize three independent realizations of the  $40 \,h^{-1} {\rm Mpc}$ comoving (cMpc) simulation boxes, each offering a spatial resolution of $100 \dim{pc}$ in proper units. This setup enables the accurate modeling of the IGM properties and yields ample independent lines of sight data for our analyses. By adopting different `DC mode' values \citep{gnedin11} for different independent realizations, the CROC simulations model varied reionization histories in separate simulation boxes, which allows us to investigate the impact of reionization history on our findings.

\begin{figure}[ht]
\centering
\includegraphics[width=\columnwidth]{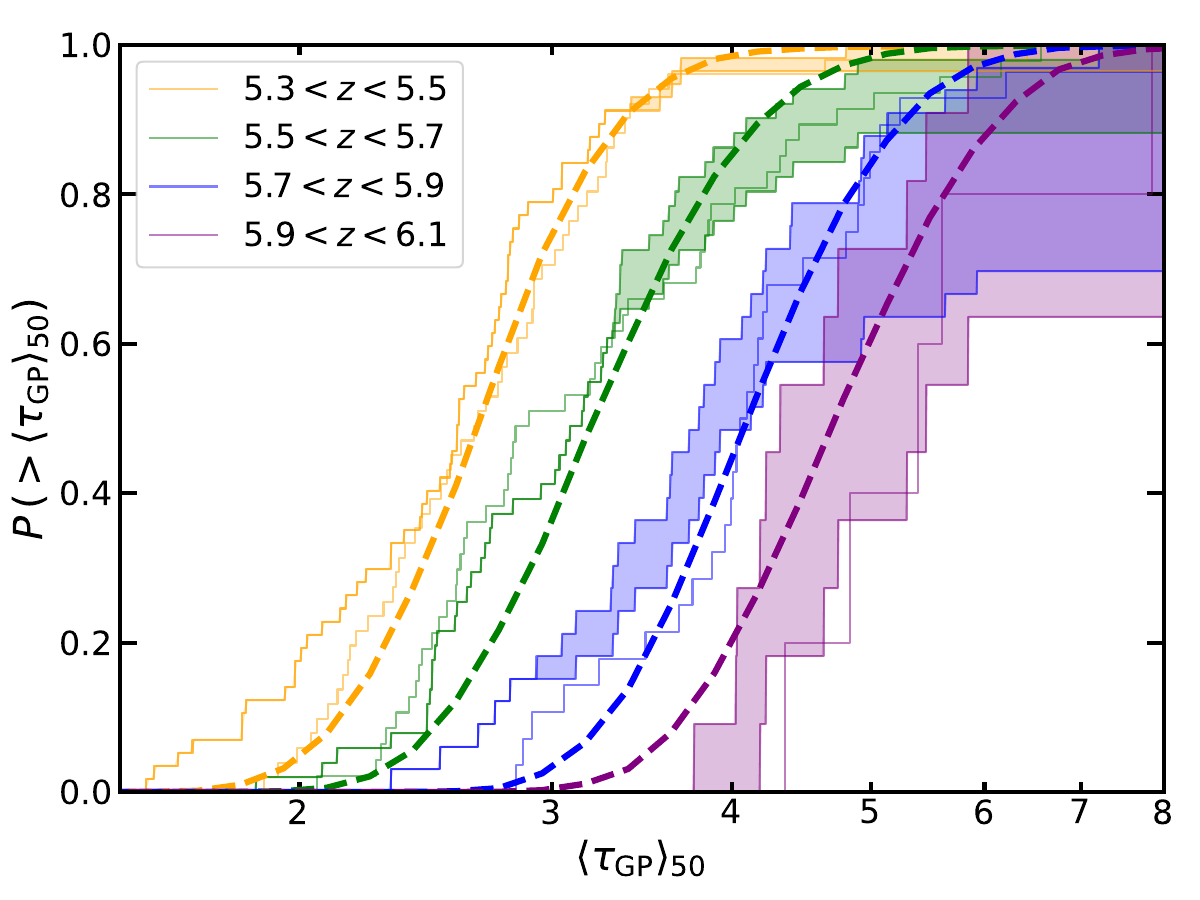} 
\caption{Distribution of mean opacities in $50h^{-1}\dim{cMpc}$ LOSs for the combination of the three CROC simulations (thick dashed lines). Thin solid lines and bands show the observational data from \citet{Becker2015} and \citet{Bosman2018}. The three CROC simulations together marginally match the distribution of opacities (at least at $z>5.5$).\label{fig:ptau}}
\end{figure}

In this paper, we use three different CROC simulations that we label ``early", ``intermediate" and ``late reionization". These three simulations sample the full range of reionization histories from 6 independent random realizations of initial conditions. Hence, the ``late" and ``early" models roughly correspond to $\pm1\sigma$ spread in possible reionization histories, while the ``intermediate reionization" history is close to the cosmic mean. In Figure~\ref{fig:ptau}, we show the distribution of mean opacities obtained from $50h^{-1}\dim{cMpc}$ LOSs from all three simulations and compare to observational data. Collectively, these simulations provide a marginal fit to the observational data - unfortunately, no other reionization simulation currently offers a better fit. The CROC simulations also match the observed distribution of long gaps in quasar spectra \citep{Gnedin2022}, but only under the ``late reionization" model. In other words, while neither CROC nor any other existing reionization simulation can match all the observational data, some of the CROC boxes do meet certain observational constraints. Hence, we have some confidence that CROC captures the key physical processes, justifying our use of these simulations despite known limitations in how precisely we capture all details of reionization.

\subsection{Synthetic \Lya\ Absorption Spectra}\label{sec:meth:synthetic_spec}

In comoving space, \Lya\ optical depth is obtained by integrating along the line-of-sight:
\begin{equation} \label{eqn:tau}
\tau(\lambda)=\sigma_0 \int n_{\rm HI}(x) \frac{c}{\sqrt{\pi} b_x} e^{-\frac{\left(u_\lambda-u_x\right)^2}{b_x^2}} \frac{d x}{1+z_x},
\end{equation}
where $\sigma_0$ is the cross section, $n_{\rm HI}$ is the HI number density, $c$ is the speed of light, $b_x$ is the Doppler parameter at position $x$ along the line-of-sight, $u_\lambda$ is the velocity corresponding to the observed wavelength $\lambda$, $u_x$ is the gas velocity at $x$, and $z_x$ is the redshift at $x$. Then the flux is
\begin{equation} \label{eqn:flux}
F(\lambda) = \exp(-\tau(\lambda))
\end{equation}
We note that $u_\lambda$ and $u_x$ include both the Hubble flow and gas peculiar velocity. 

We adopt two critical simplifications in this first study. Peculiar velocities induce a surjective but non-injective mapping from comoving space to velocity space, thereby obscuring the direct association between gas properties (sampled in comoving space) and observed flux (in velocity space). To avoid this additional complication and focus on the fundamental physical connection between gas properties and transmission spikes, our first simplification is to use synthetic \Lya\ absorption spectra in ``real space", i.e.\ assuming a zero peculiar velocity ($v_{pec} = 0$). This approach is equivalent to using the ``Fluctuating Gunn-Peterson Approximation" \citep{fgpa0,fgpa1,fgpa2}, which has been instrumental in understanding the Ly$\alpha$ forest at lower redshifts. While it is crucial to account for peculiar velocities for interpreting real observational data, our current focus is on the theoretical investigation of the physical causes behind transmission spikes. Therefore, we defer the inclusion of peculiar velocities to follow-up works.

Our second simplification is using a Doppler profile instead of the correct Voigt profile for convolving spectra. This choice effectively excludes non-local effects from high-column density damped Lyman-alpha systems (DLAs). The Lorentzian wings of a DLA can impact absorption far from its actual location in velocity space, thereby obscuring the relationship between the transmitted flux and the physical properties of the IGM. Similarly, applying the correct profile is essential for correctly interpreting observational data, a step we plan for future work.

As our data sample, we generate 100{,}000 line-of-sights (LOS) at each redshift, randomly oriented to sample the full volume of the simulation boxes. We note that the results shown in this paper converge when using only a tenth of these data. Along these LOS, we record the \Lya\ flux and the physical properties of the gas, as well as the intensity of the radiation field. We provide the details in Section~\ref{sec:sub:predict}. 

\subsection{Explainable Boosting Machines}\label{sec:meth:ebm}
EBMs provide a fitted relationship between the target quantity $y$ and the features $\vtheta \in \mathbb{R}^{n}$. They are specifically designed to be \emph{interpretable}, i.e.\ the dependence of the target quantity on the features is given in explicit functional forms. Mathematically, we have
\begin{equation} \label{eqn:ebm}
\ebmy = \betay + \sum_{i=0}^{\np-1} \fyi(\theta_i) + \sum_{i=0,i \neq j}^{\np-1}\sum_{j=0}^{\np-1}  \fyij(\theta_i, \theta_j)
\end{equation}
where $\ebmy$ is the predicted value of the target quantity $y$ given $\np$ features $\vtheta$, $\betay$ is the baseline (c.f.\ the mean) value of the target quantity $y$, and $\fyi$ and $\fyij$ are piecewise one- and two-dimensional functions, respectively. The magnitudes of $\fyi$ and $\fyij$ indicate the relative importance of each feature in predicting the target quantity. We refer to $\fyi$ as feature functions and $\fyij$ as interaction functions. An example of employing EBMs in astrophysics is presented in \citet{Hausen2023}, where EBM models reveal the relative importance of different dark matter halo properties in setting galaxy stellar mass and star formation rate.

We evaluate the EBM performance by computing the $r^2$ variance metric:
\begin{equation}
r^2=1-\frac{\sum_{i=0}^{N-1}\left(y_i-\hat{y}_i\right)^2}{\sum_{i=0}^{N-1}\left(y_i-\bar{y}\right)^2},
\end{equation}
where $N$ is the number of objects, $y_i$ is the true value of the target quantity for object $i$, and $\hat{y}_i$ is the predicted value from the model for object $i$. $r^2$ measures the extent to which the variance in the actual outcomes is captured by the model. A higher $r^2$ value indicates that the model's predictions more closely match the actual data.

In terms of the training procedure, we use the InterpretML \citep{nori2019} implementation of EBMs, and adopt most of the default hyperparameter values for InterpretML version 0.4.4., except for ``max bins'', which we set to 4096, after performing a grid search for optimal model performance using the allocated subset of data for training. 

We adopt a standard 80/20 train-test split, allocating 80\% of our dataset for training the models and the remaining 20\% for assessing their performance.

\section{Results}\label{sec:results}
\subsection{Predictive Power of Models}\label{sec:sub:predict}

Our first objective is to train the EBM models to accurately predict the \Lya\ flux. Given that EBMs reveal the relative contribution of each input on the prediction outcome, we can quantify the importance of individual physical properties in generating the spectra. We first investigate a range of physical properties as inputs for training our EBM models based on the physical relationships between the inputs.  In Section~\ref{sec:sub:contribution}, we use EBM to explicitly quantify the relative importance of each feature.

\begin{figure*}
\centering
\includegraphics[width=\textwidth]{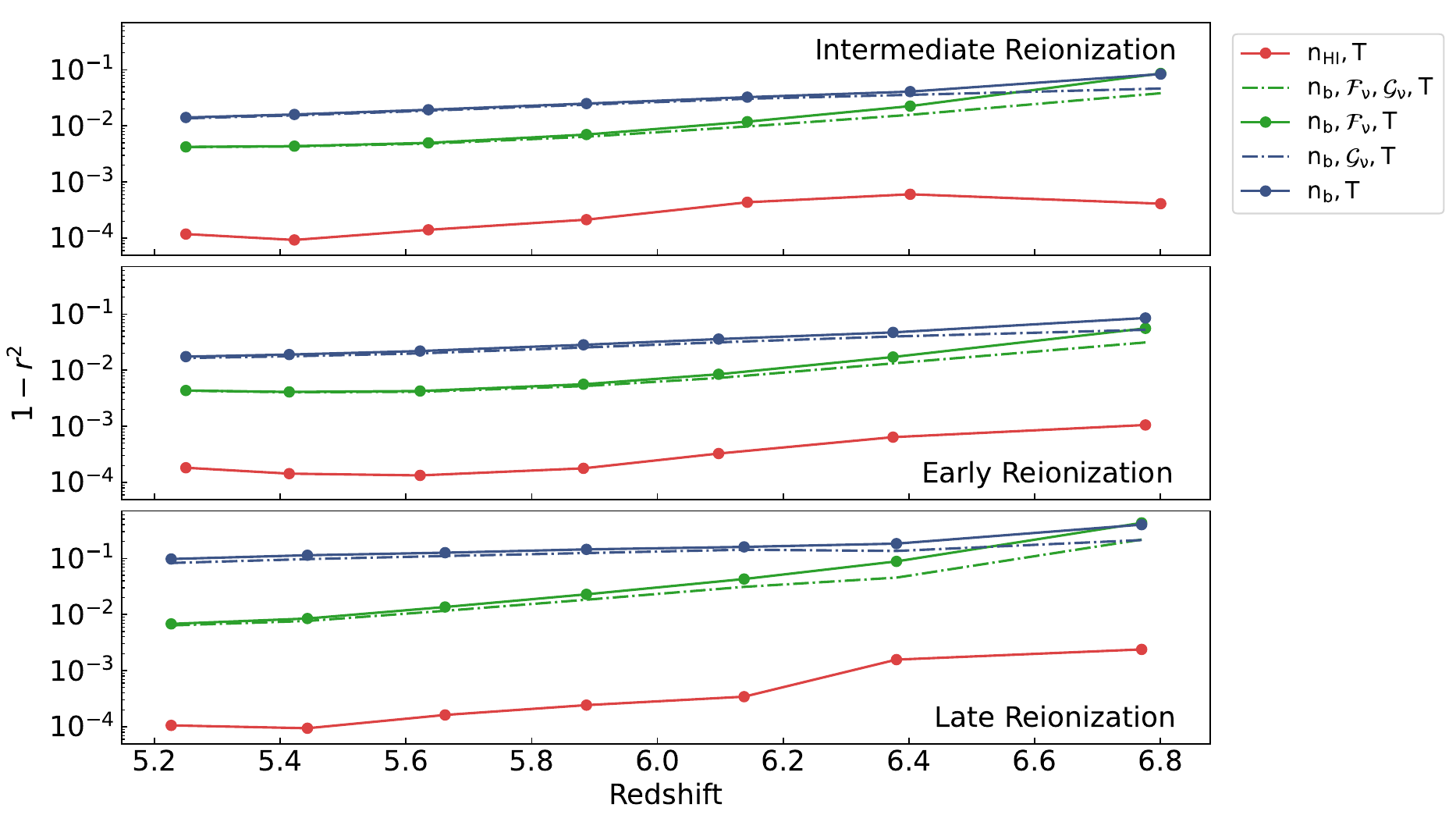}
\caption{EBM model performance, measured using the $1-r^2$ ``missing variance" metric, as a function of redshift for models trained with different input combinations.  Each panel corresponds to a different reionization history. As a baseline, with HI number density ($n_{\rm HI}$) and temperature ($T$) as inputs, the model predictions are highly accurate. Then $n_{\rm HI}$ is replaced by combinations of baryon number density ($n_b$) with ionization parameters, $\fnu$ (representing the cosmic background) and $\gnu$ (representing the local ionizing sources). The results indicate that $\fnu$ is the predominant component of ionizing radiation.}
\label{fig:model_performance}
\end{figure*}

At first glance, Equations~\ref{eqn:tau} and \ref{eqn:flux} suggest a direct dependence of flux on HI number density ($n_{\rm HI}$) and temperature ($T$), as the Doppler parameter is a function of $T$. Consequently, we anticipate that using $n_{\rm HI}$ and $T$ as inputs for the EBMs will yield a highly accurate prediction of the flux. We convolve $n_{\rm HI}$ along the line-of-sight using the Gaussian profile $e^{-x^2/b_x^2}$, where $b_x(T)$ is the temperature-dependent Doppler width. This convolution accounts for local temperature variations, employing a Gaussian profile specific to the temperature at each point along the line-of-sight. In Figure~\ref{fig:model_performance}, we show the $1-r^2$ values on a logarithmic scale as a function of redshift. Uncertainties associated with these values, obtained after training with three different sets of LOSs, are comparable to the width of the line on the plot for $z \leq 6.4$, and slightly larger than the width of the plotted line beyond this redshift. The three panels represent different boxes with different reionization histories. We first only focus on red lines, which show the model predictions with $\{ n_{\rm HI}, T\}$ as inputs. Consistent with our expectations, in all three boxes, the model predictions with $\{ n_{\rm HI}, T\}$ inputs are the best performing. A $1-r^2$ value of $10^{-3}$ means that 0.1\% of the total variance in the target quantity is not captured by the model. We find larger discrepancies between our predictions and the test data primarily at the peak regions of transmission spikes. These discrepancies stem from the limitations of the EBM algorithm, which models the target quantity using piecewise constant functions and thus loses accuracy in regions with sparse data points.

\subsection{Role of Ionizing Radiation}\label{sec:sub:radiation}

The HI number density is affected by the baryon number density ($n_{\rm b}$) and the prevailing ionizing radiation in the universe. Similar to what we did to $n_{\rm HI}$, we also convolve $n_{\rm b}$ along the line-of-sight, applying a temperature-dependent Gaussian profile that varies based on the temperature at each point. As a baseline, we first use the baryon number density ($n_{\rm b}$) and temperature ($T$) as inputs. The light blue lines in Figure~1 represent models trained using Gaussian convolved $\{n_{\rm b}, T\}$ inputs. The quality of the predictions is much lower compared to the $\{n_{\rm HI}, T\}$ input case. This difference in $r^2$ values illustrates the important role of ionizing radiation in producing the \Lya\ flux. This leads to the question: \textit{How can we quantify ionizing radiation at each point along the line-of-sight?} In the CROC simulations, the radiative transfer is implemented using the OTVET method \citep{gnedin14}. To summarize, in the simulations, the radiation energy density $E_\nu$ is computed as
\begin{equation}
E_\nu=\bar{E}_\nu \fnu + (\gnu-\bar{\gnu} \fnu),
\end{equation}
where $\bar{E}_\nu$, a constant, is the spatial average (i.e.\ the cosmic background). The term $\gnu$ accounts for the local ionizing sources inside the simulation box, and can be written as the sum of fluxes from all sources inside the box,
\[
    \gnu(\vec{x}) = \frac{1}{4\pi c a^2} \sum_i \frac{L_{i,\nu}}{|\vec{x}-\vec{x}_i|^2} e^{-\tau_\nu(\vec{x},\vec{x}_i)},
\]
where $\vec{x}_i$ and $L_{i,\nu}$ are the comoving location and the luminosity of source $i$ inside the simulation box, and $\tau_\nu(\vec{x},\vec{x}_i)$ is the optical depth between the spatial locations $\vec{x}$ and $\vec{x}_i$.

$\fnu(\vec{x})$ is the angle average of $f_\nu(\vec{x},\vec{n})$ that satisfies the following equation:
\begin{equation}
    \frac{a}{c}\frac{\partial f_\nu}{\partial t} + \vec{n}\frac{\partial f_\nu}{\partial \vec{x}} = -k_\nu f_\nu + f_\nu \langle k_\nu f_\nu\rangle,
    \label{eq:fnu}
\end{equation}
where $k_\nu(\vec{x})$ is the absorption coefficient and 
\begin{equation}
    \fnu(\vec{x}) \equiv \frac{1}{4\pi} \int d\Omega f_\nu(\vec{x},\vec{n}).
    \label{eq:favg}
\end{equation}
Physically, $\fnu-1$ can be interpreted as the fluctuation in the ``far radiation'' (i.e.\ radiation outside the simulation volume), and $\langle\fnu\rangle=1$.

To account for ionizing radiation in our model, we have included $\fnu$ and $\gnu$ in the input parameters. This is depicted in Figure~\ref{fig:model_performance}, where the $\{\fnu, \gnu, n_{\rm b}, T\}$ input case is represented by dark green lines. To separate the contribution from $\fnu$ and $\gnu$, we also use  $\{\fnu, n_{\rm b}, T\}$ and $\{\gnu, n_{\rm b}, T\}$ as inputs to train the models, illustrated in light green and dark blue lines, respectively. Our analysis indicates that $\fnu$ is the predominant component of ionizing radiation; $\gnu$ moderately improves model performance at early times. This finding is significant, as previous studies primarily linked transmission spikes to local ionizing sources such as galaxies. In contrast, our study, which integrates numerical simulations with an interpretable machine learning algorithm, suggests that density is a key factor in generating these spikes, and that background radiation plays a more crucial role than local ionizing sources. We also note that as redshift increases, the performance of the EBM models decreases. This trend is linked to the increasing IGM opacity at higher redshifts, which makes transmission spikes less frequent and requires a larger number of sightlines for effective model training.  However, we are already using more sightlines than could be observed even with future 30-meter class telescopes. We therefore limit this analysis to our existing number of sightlines to focus on physical interpretation instead of generating more for the sake of improved model performance.

Additionally, we note variations in the performance of the $\{n_b, ...\}$ models trained and evaluated on data from boxes with different reionization histories. The models based on the intermediate and early reionization boxes yield comparable results across the entire redshift range considered.  However, the late reionization box shows relatively worse model performance, particularly at higher redshifts. This trend is consistent with our expectations: in the late reionization scenario, particularly at $z>6$, the universe is still going through rapid reionization, presenting a more complex environment for EBMs to accurately model. Regardless of reionization scenario, the effects of the cosmic background radiation on \Lya\ transmission peaks are dominant to those from local ionizing sources.

\subsection{Non-equilibrium Effects}\label{sec:sub:sub:non-equlibrium}

In theory, the three physical quantities $\fnu$, $\gnu$, and $n_{\rm b}$ should collectively determine $n_{\rm HI}$. Therefore, we would expect a model trained with $\{\fnu, \gnu, n_{\rm b}, T\}$ to have the same level of performance as one trained with $\{n_{\rm HI}, T\}$ inputs. However, as shown in Figure~\ref{fig:model_performance}, the $\{\fnu, \gnu, n_{\rm b}, T\}$ model consistently exhibits lower $r^2$ values across all redshifts. Moreover, the ratio of $r^2$ values between the $\{\fnu, \gnu, n_{\rm b}, T\}$ model and $\{n_{\rm HI}, T\}$ model decreases with increasing redshift, indicating a decline in the $\{\fnu, \gnu, n_{\rm b}, T\}$ model performance at higher redshifts.

We posit that the decrease in model prediction power is due to non-equilibrium effects that the EBM model cannot capture. Under the assumption of ionization equilibrium, we can relate $n_{\rm HI}$ to $n_{\rm b}$ using the following equation:
\begin{equation} \label{eqn:ionization_eq}
   n_{\HI}^{\rm eq} = \frac{R(T)n_e n_{\HII}}{\Gamma},
\end{equation}
where $R(T)$ is the recombination coefficient, $n_e$ ($\propto n_b$) is the electron number density, $n_{\rm HII}$ is the HII number density, and $\Gamma$ is the photo-ionization rate. However, the straightforward mathematical relation does not hold when residual non-equilibrium effects from cosmic reionization influence the neutral hydrogen number density. When the relationship between $n_{\rm HI}$ and $n_e$ does not simply follow Equation~\ref{eqn:ionization_eq}, the accuracy of the EBM in mapping $n_b$, $\fnu$ and $\gnu$ to $n_{\rm HI}$ decreases. This argument is consistent with our finding that $\{\fnu, \gnu, n_{\rm b}, T\}$ model performance decreases when the correlation coefficient between $n_{\rm HI}$ and $n_{\HI}^{\rm eq}$ is smaller.

\begin{figure}[ht]
\centering
\includegraphics[width=\columnwidth]{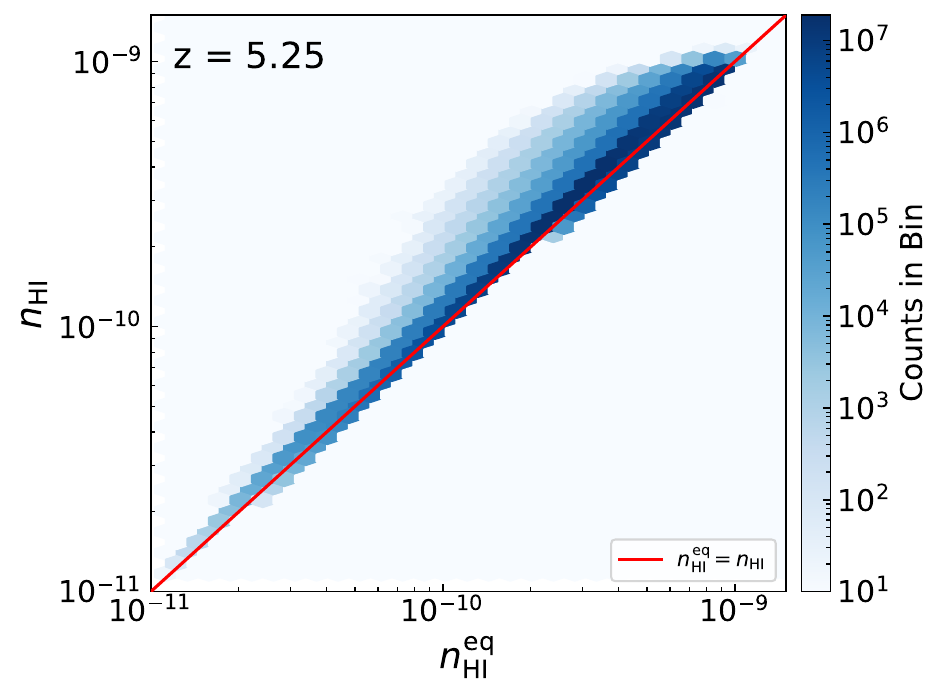}
\caption{$n_{\rm HI}$, sampled along line-of-sights versus $n_{\HI}^{\rm eq}$, calculated with Equation~\ref{eqn:ionization_eq} assuming ionization equilibrium, at z=5.25. Colors show the number counts in each hexbin. We observe significant deviations of many data points from the equilibrium line.}\label{fig:non_eq}
\end{figure}

\begin{figure}[ht]
\centering
\includegraphics[width=\columnwidth]{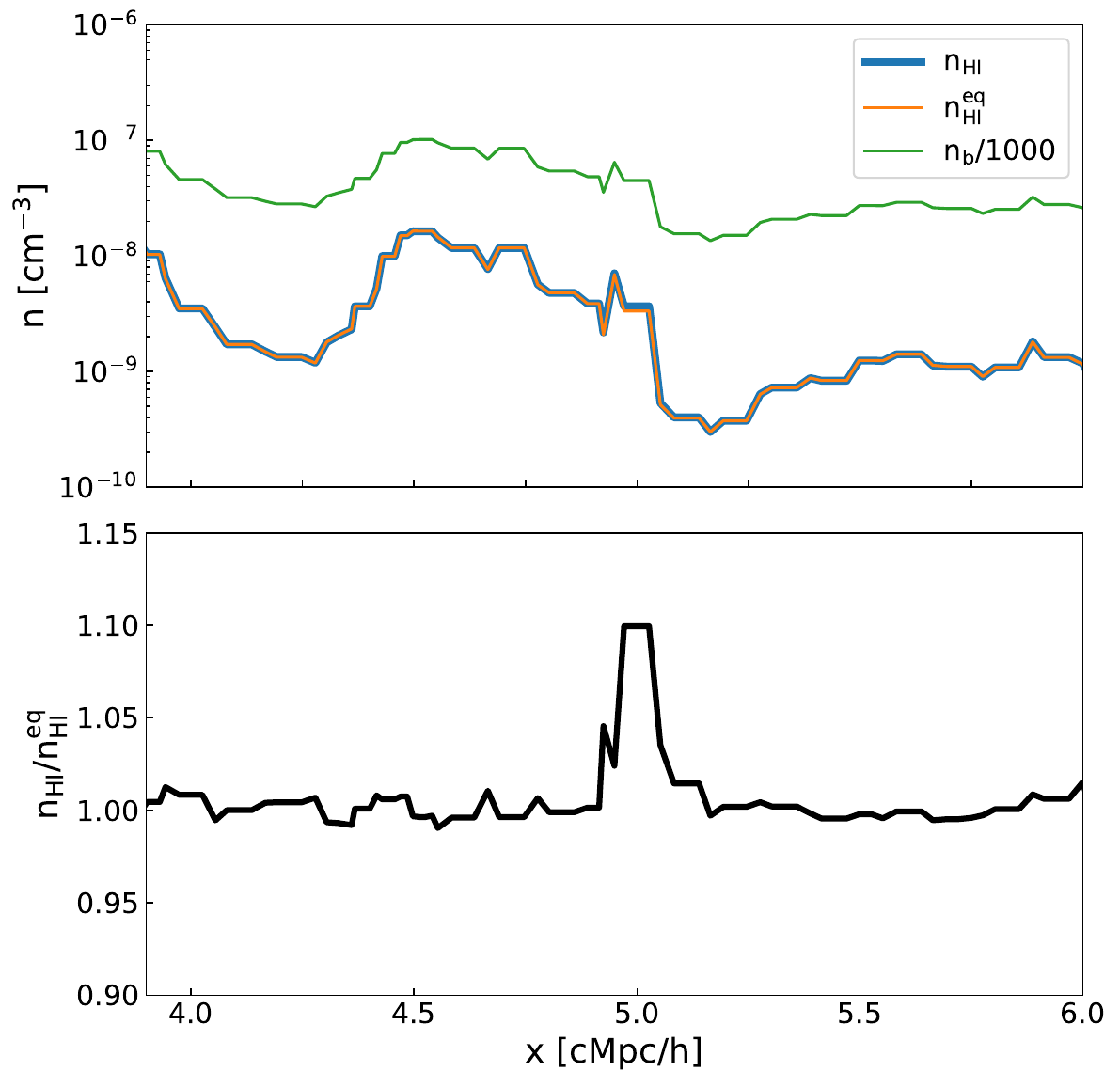} 
\caption{An example of non-equilibrium effects along a random line-of-sight. Top: $n_{\rm HI}$, $n_{\HI}^{\rm eq}$ calculated assuming ionization equilibrium and rescaled $n_{b}$ as a function of LOS position. Bottom: Ratio of $n_{\rm HI}$ and $n_{\HI}^{\rm eq}$ as a function of LOS position. A noticeable deviation from equilibrium coincides with a drop in $n_{b}$, suggesting the presence of a shock at this location.}\label{fig:losneq}
\end{figure}

We demonstrate the non-equilibrium effects in Figure~\ref{fig:non_eq}, which shows the simulated $n_{\rm HI}$ versus $n_{\HI}^{\rm eq}$ computed using Equation~\ref{eqn:ionization_eq}. There is a notable deviation in many data points from the equilibrium line. Figure \ref{fig:losneq} presents an example of non-equilibrium effects along a random line of sight. We observe a distinct deviation from equilibrium, highlighted by the ratio of $n_{\rm HI}$ and  $n_{\HI}^{\rm eq}$ displayed in the bottom panel of Figure~\ref{fig:losneq}, which coincides with a drop in $n_b$. This pattern suggests the possible presence of a shock at this location. We expect non-equilibrium effects to be important in shock regions; as the temperature jumps across the shock, it takes time for the neutral fraction to decrease to a new equilibrium value, corresponding to the lower recombination rate at a higher temperature.

A similar comparison at other redshifts does not show any clear redshift trend. At higher redshifts the reionization is incomplete, so one might expect non-equilibrium effects to become relatively more important with increasing redshift. On the other hand, shocks are stronger at lower redshifts as larger wavelengths become non-linear, so there is also an argument for the importance of the non-equilibrium effects to increase with decreasing redshift. In fact, the behavior of the red curve in Fig.\ \ref{fig:model_performance} appears to show that both trends take place.

\subsection{Contribution of Each Physical Quantity}\label{sec:sub:contribution}

\begin{figure*}[t]
\centering
\includegraphics[width=\textwidth]{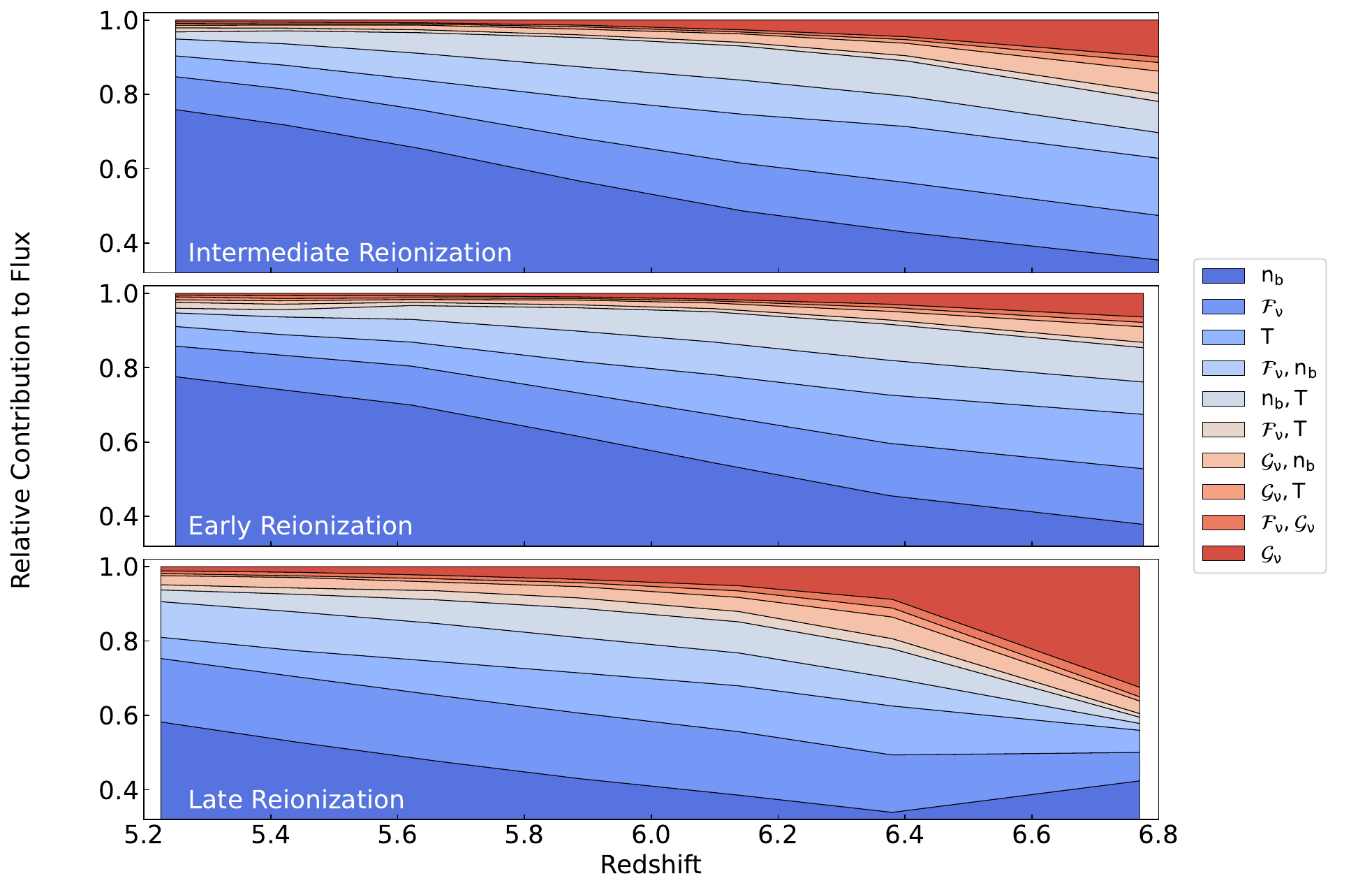}
\caption{Normalized average contributions (scaled to sum to one) of IGM properties to \Lya\ flux prediction. Each panel corresponds to a different reionization history. In intermediate and early reionization scenarios, the contribution of $n_b$ decreases with increasing redshift, while the impact of $\gnu$ grows, aligning with the expected greater influence of local ionization at higher redshifts. A similar, yet non-monotonic trend is observed in the late reionization scenario, indicative of a universe still undergoing rapid reionization.}
\label{fig:feature_contribution}
\end{figure*}

As discussed in Section~\ref{sec:meth:ebm}, the univariate and bivariate functions $\fyi$ and $\fyij$ from Equation~\ref{eqn:ebm} quantify the contribution of each feature. We can then define a summary statistic, the average contribution of each feature (denoted as $\bar{f}_y^i$) such that 
\begin{equation}
    \label{eqn:average_contribution}
    \bar{f}_y^i =  \frac{\sum_{j=0}^{N_{bin}-1} |f(\theta_{i,j})|N_j}{\sum_{j=0}^{N_{bin}-1}N_j},
\end{equation}
\noindent
where $f$ represents the feature or interaction function ($\fyi$ or $\fyij$), $\theta_{i,j}$ the value of feature $\theta_i$
in the $j$th bin, $N_j$ the number of samples in bin $j$, and $N_{bin}$ the total number of bins. 

In Figure~\ref{fig:feature_contribution}, we revisit the $\{\fnu, \gnu, n_{\rm b}, T\}$ input case and show the normalized average contributions (scaled to sum to one) across three simulation boxes with different reionization histories. Consistently across all redshifts, $n_b$ emerges as a dominant factor in predicting the \Lya\ flux. In boxes of intermediate and early reionization, the contribution of $n_b$ decreases with increasing redshift, while the influence of $\gnu$ (the local sources) increases. This pattern is consistent with the expectation that ionizing radiation from local sources has a greater impact at higher redshifts when the universe is still undergoing substantial reionization. In the late reionization scenario, a similar but non-monotonic trend is evident, reflecting phenomena (e.g. overlapping ionizing bubbles) characteristic of a universe still undergoing rapid reionization.

The specific forms of functions $\fyi$ and $\fyij$ are shown in Appendix \ref{sec:app}.

\subsection{Interpreting Observations}\label{sec:sub:obs}

Several observational studies have measured the average \Lya\ transmission flux as a function of the distance from nearby galaxies \citep[e.g.][]{Kakiichi2018, Meyer2019,Meyer2020,Kashino2023,Christenson2023}. Overall, observations paint a complex and sometimes contradictory picture. The most clearly established trend across all redshifts $z>5$ is the decrease in the transmitted flux within about 1 pMpc from galaxy locations. The same trend is found in CROC \citep{Garaldi2019} and another similar large-scale simulation project, ``THESAN" \citep{Garaldi2022}.

\begin{figure*}[t]
\centering
\includegraphics[width=\textwidth]{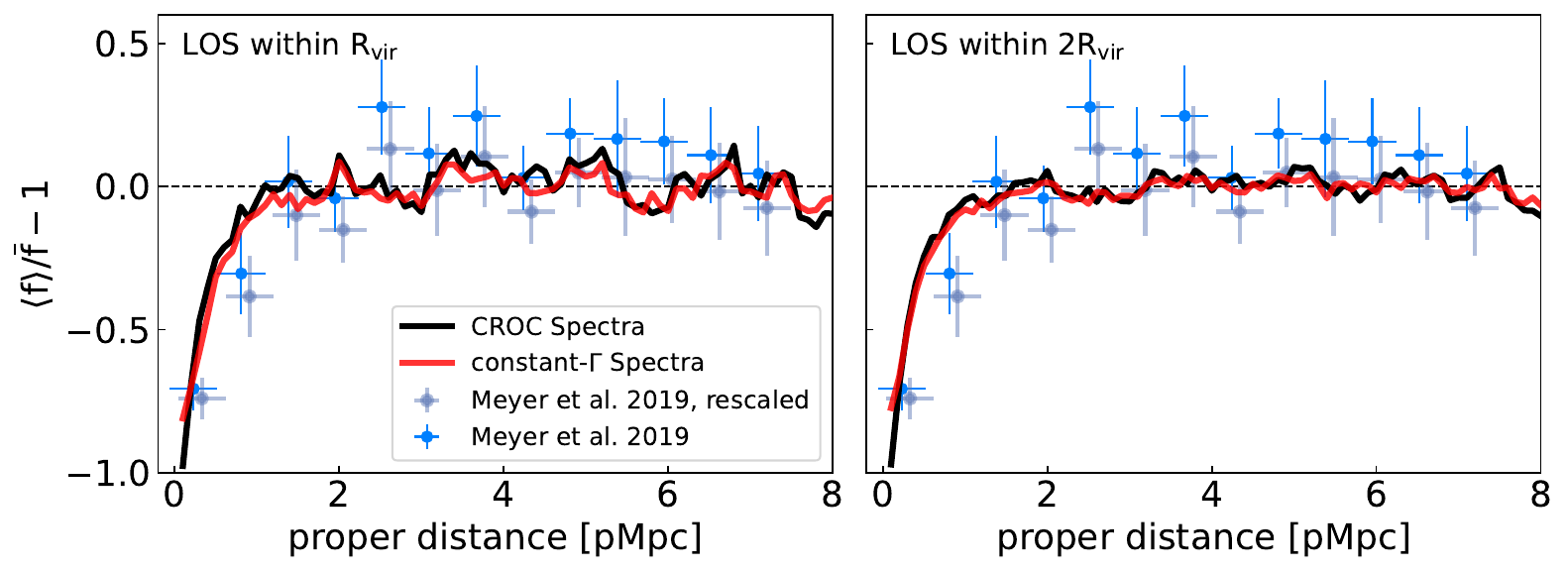}
\caption{Normalized average transmitted flux as a function of distance from galaxies. We show measurements using both the simulated spectra and spectra generated assuming a spatially uniform photo-ionization rate ($\Gamma$). Points with error bars reference measurements from \citet{Meyer2019}; brighter blue points show the original measurement and lighter blue points show the measurement with the rescaled mean flux (see text for details). The observed flux suppression at small distances in both datasets suggests that galaxies are not the primary cause, since galaxies influence IGM properties through $\Gamma$. }\label{fig:constant_gamma}
\end{figure*}

Taken at face value, the observed decrease in the transmitted flux is inconsistent with our finding that the radiation from nearby galaxies has a negligible impact on the transmitted flux at the late stages of cosmic reionization. In order to understand this apparent discrepancy, we also generate \Lya\ transmission spectra assuming a spatially uniform photo-ionization rate ($\Gamma$), using the global mean of each box at every redshift. These spectra allow us to control for local ionizing sources. Since galaxies influence IGM properties through photo-ionization, the constant-$\Gamma$ spectra effectively remove the impact of galaxies on the IGM. If flux suppression still occurs at small distances in our control sample, it would indicate that galaxies are not the primary cause.

Figure~\ref{fig:constant_gamma} shows our measurement of the average \Lya\ transmission flux as a function of distance to galaxies, comparing simulated \Lya\ transmission spectra (in black) with spectra generated assuming a constant $\Gamma$ (in red). The left panel considers galaxies only if their virial radii ($R_{vir}$) exceed the radial distance from the dark matter halo center to the LOS, whereas in the right panel, we relax this criterion to $2 \times R_{vir}$. The black and red lines in Figure~\ref{fig:constant_gamma} show the same level of anti-correlation at distances less than 2 pMpc. Hence, in the simulations, the observed flux suppression is entirely due to the large-scale correlation between halos and density, and galaxies simply serve as biased tracers of the large-scale structure.

For illustration, we also show the observed points from \citet{Meyer2019}. The observations appear to be offset from zero for all distances above 2 pMpc, which may indicate a potential bias in the observationally determined value for $\bar{f}$. Such a bias in observations is possible, since \citet{Meyer2019} compute $\bar{f}$ as the mean flux within 7.5 pMpc around each galaxy, and at such small distances the halo-density correlation remains non-negligible.  In fact, the average halo-mass bias over the distance of 7.5 pMpc at $z=5.4$ is 0.14, 0.074, and 0.045 for halos of mass $10^{12}$, $10^{11}$, and $10^{10}$ solar masses, respectively. To account for the potential bias, we also rescale $\bar{f}$ by 13\% to show how the observational data would appear if the mean flux in the observations were computed differently. The correspondence between rescaled observational data and the CROC data lends support to our proposition that the observed flux suppression is entirely due to the halo-density correlation.

\section{Summary and Discussion} \label{sec:discussion}

\begin{figure}[ht]
\centering
\includegraphics[width=\columnwidth]{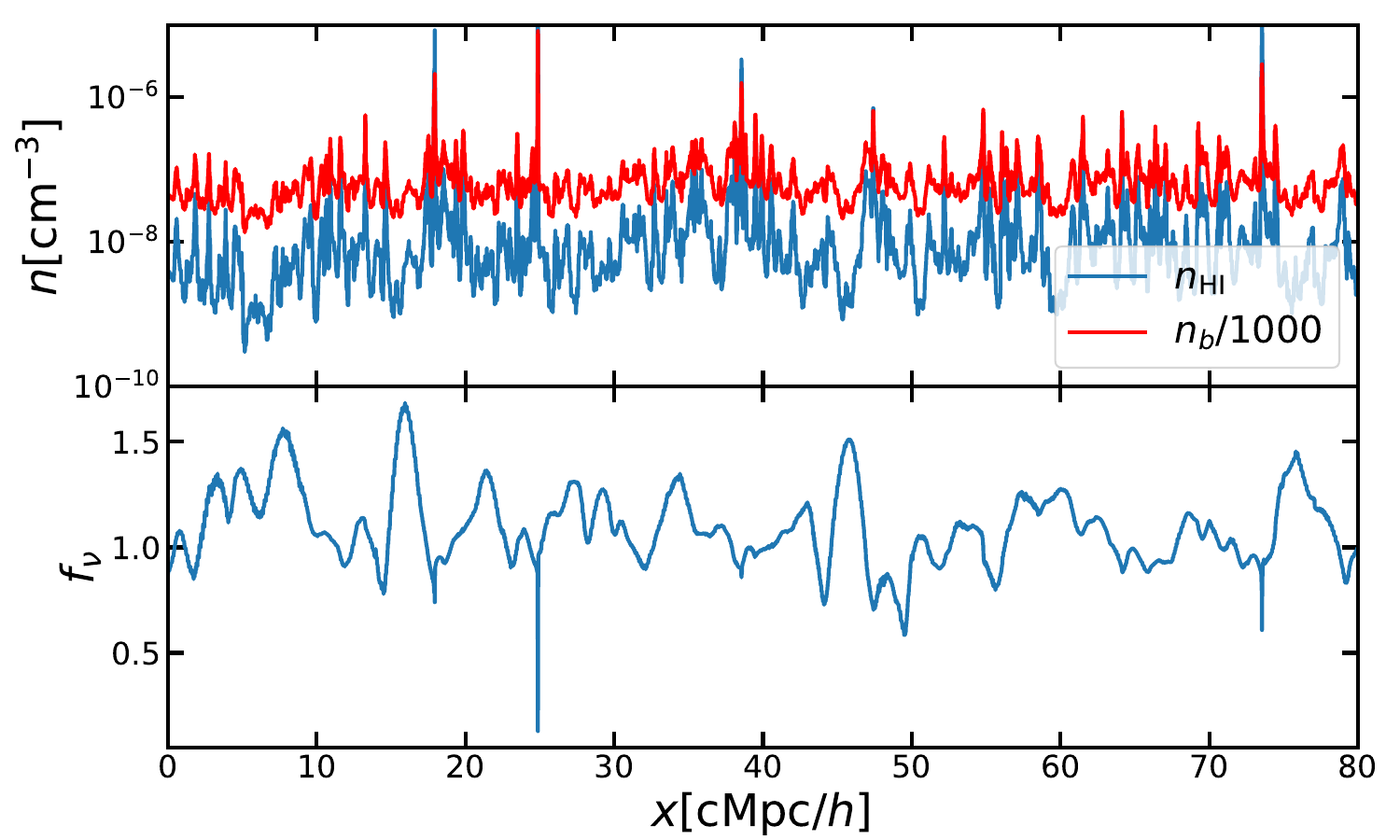} 
\caption{Baryon and HI number densities (top) and $\fnu$ (bottom) along a random line of sight. $\fnu$ shows an anti-correlation with density, albeit not a strong one. \label{fig:losf}}
\end{figure}

There are two main conclusions from this work. First, somewhat unexpectedly, we find that radiation from nearby galaxies plays a negligible role in controlling the transmitted flux in quasar absorption spectra. The observed strong correlation between the transmitted flux and galaxy locations is explained entirely by the correlation of the transmitted flux with cosmic large-scale density distribution, of which galaxies are just a biased tracer.

Second, we find that in addition to the obvious factor determining the transmitted flux - the local gas density, the flux is also affected by a component of the radiation field that describes ``cosmic background", $\fnu$ from Equations \ref{eq:fnu} and \ref{eq:favg}. The complete physical interpretation of that result is elusive and will require substantial additional effort, and we delegate it to future work. What can be said immediately is that $\fnu$ does not depend on the local distribution of sources (Equation~\ref{eq:fnu}) but does depend on the large-scale cosmic density (via the absorption coefficient $k_\nu$), and hence encodes information about local variations in the photon free path (of which the commonly known photon mean free path is the mean). Figure \ref{fig:losf} shows the baryon density, the neutral hydrogen density, and $\fnu$ along a random line of sight. While some anti-correlation of $\fnu$ with density smoothed on $\sim 1$ Mpc scale is apparent from the figure, the two are not equivalent. 

In fact, we find no smoothing scale $R_S$ such that the baryon density smoothed on scale $R_S$ along the line of sight (i.e.\ in 1D) approximates $\fnu$. It might be possible to find a smoothing scale for the 3D baryon density that offers a better match to $\fnu$. We leave such an investigation to future work since a more complex and laborious exploration may be required to come up with a better physical interpretation of $\fnu$. At this point, it is sufficient for our purpose that we can present an equation for $\fnu$ and a plausible physical interpretation as a variation in the photon free path.

\acknowledgments

This work was supported in part by the NASA Theoretical and Computational Astrophysics Network (TCAN) grant 80NSSC21K0271. This manuscript has also been co-authored by Fermi Research Alliance, LLC under Contract No. DE-AC02-07CH11359 with the United States Department of Energy. This work used resources of the Argonne Leadership Computing Facility, which is a DOE Office of Science User Facility supported under Contract DE-AC02-06CH11357. An award of computer time was provided by the Innovative and Novel Computational Impact on Theory and Experiment (INCITE) program. This research is also part of the Blue Waters sustained-petascale computing project, which is supported by the National Science Foundation (awards OCI-0725070 and ACI-1238993) and the state of Illinois. Blue Waters is a joint effort of the University of Illinois at Urbana-Champaign and its National Center for Supercomputing Applications. CA acknowledges support from DOE grant DE-SC009193 and the Leinweber Foundation at the University of Michigan.

\appendix

\section{Univariate and Bivariate Functions}\label{sec:funcs}
\label{sec:app}

\begin{figure*}[b]
    \centering
        \includegraphics[width=\textwidth]
        {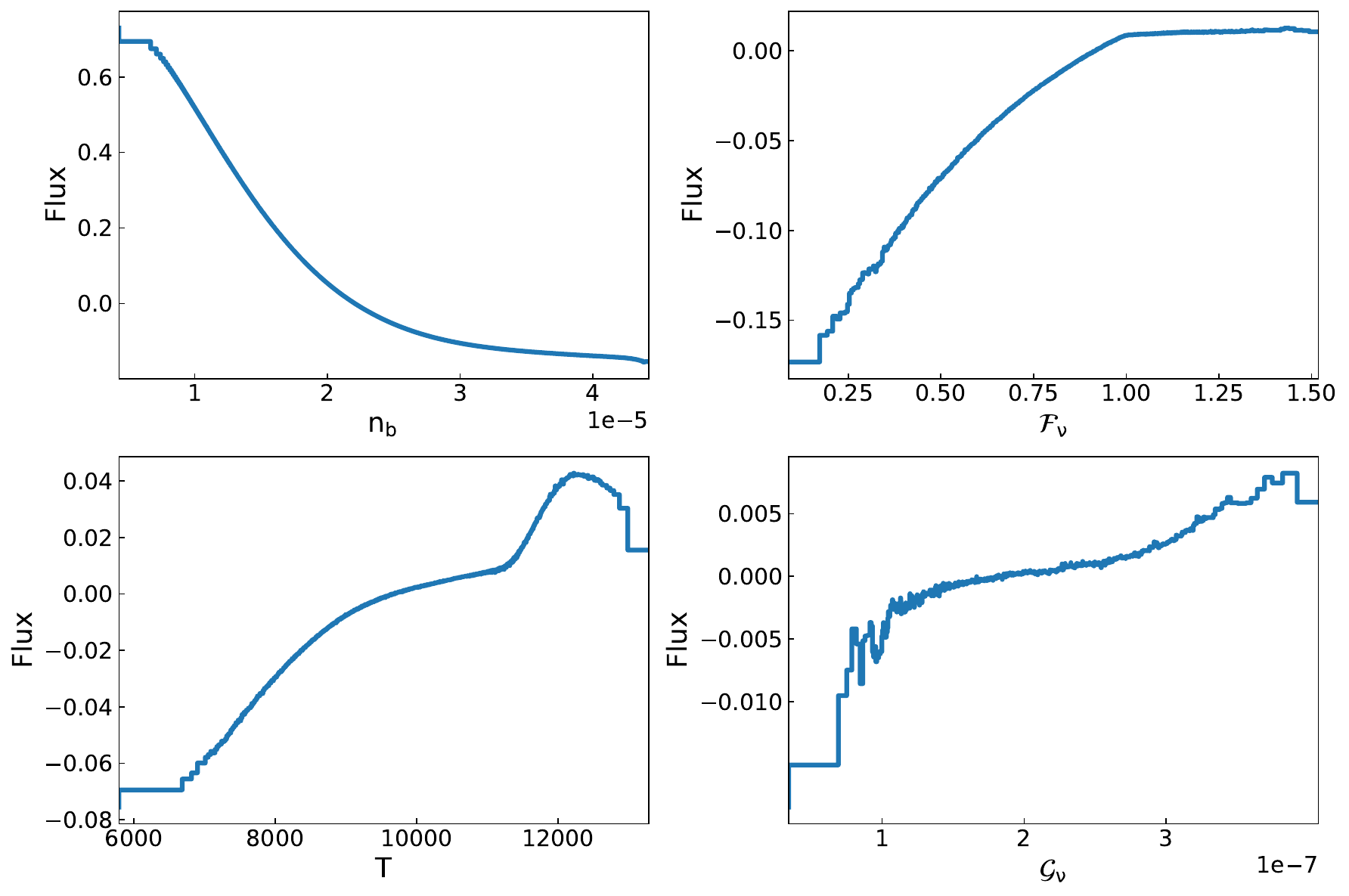}
\caption{Univariate functions at $z=5.25$ from the intermediate reionization case. There is an apparent anti-correlation between flux and $n_b$, and positive correlations between flux and $\fnu$, $T$, and $\gnu$, respectively. These trends suggest that \Lya\ transmission spikes are predominately found in regions of lower density, higher ionization, and higher temperatures. }
\label{fig:univariate}
\end{figure*}

For illustrative purposes, we use the functions at $z=5.25$ from the intermediate reionization case. The forms of the feature and interaction functions naturally vary when trained on different datasets. In Figure~\ref{fig:univariate}, we observe an apparent anti-correlation between flux and $n_b$, and positive correlations between flux and $\fnu$, $T$, and $\gnu$, respectively. These correlations paint a clear physical picture: \Lya\ transmission spikes are predominately found in regions of lower density, higher ionization, and higher temperatures. The magnitude of the flux in each panel indicates the relative importance of gas density, ionization, and temperature, in line with our previously presented results in Figure~\ref{fig:feature_contribution}.

\begin{figure*}
    \centering
        \includegraphics[width=\textwidth]
        {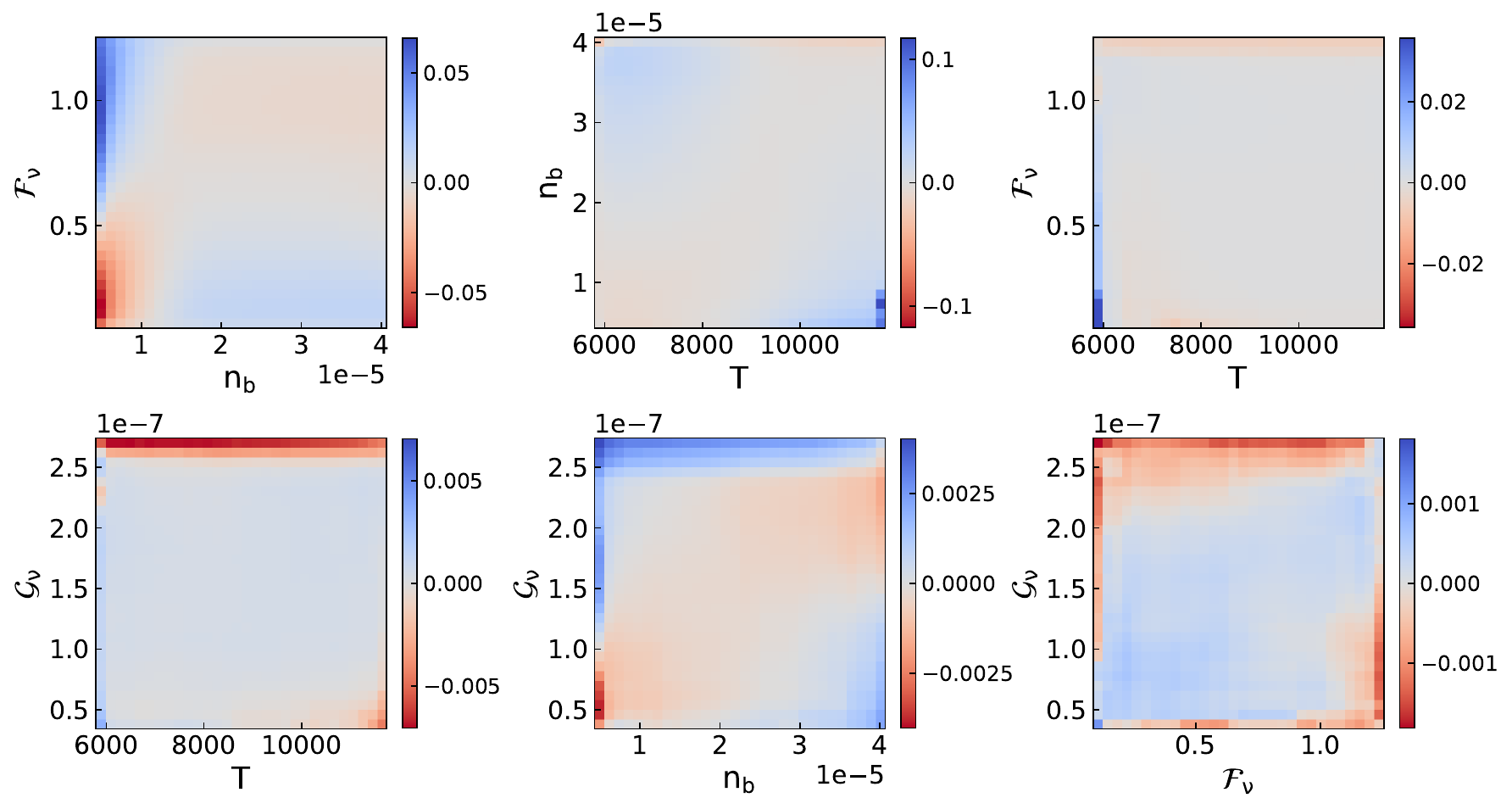}
\caption{Bivariate functions at $z=5.25$ from the intermediate reionization case.}
\label{fig:bivariate}
\end{figure*}

Additionally, we show the two-dimensional interaction functions in Figure~\ref{fig:bivariate}. We note that as shown in Figure~\ref{fig:feature_contribution}, the interaction terms are not the dominant features. Physically, in our case, the interaction terms can be thought of as second-order expansions of the first-order terms. Taking the $\fnu$ versus $n_b$ interaction in Figure~\ref{fig:bivariate} as an example, $\fnu$ (which essentially represents $\Gamma$, given the minimal contribution from $\gnu$) and $n_b$ are expected to dictate $n_{\HI}$. Referring back to Equation~\ref{eqn:ionization_eq}, we derive:
\begin{align*}
n_{\text{HI}} & \sim \frac{n^2}{\Gamma} = \frac{(\langle n \rangle + \delta n)^2}{(\langle \Gamma \rangle + \delta \Gamma)} \approx \frac{\langle n \rangle^2}{\langle \Gamma \rangle} \left( 1 + 2\frac{\delta n}{\langle n \rangle} - \frac{\delta \Gamma}{\langle \Gamma \rangle} -  2\frac{\delta n \delta \Gamma} {\langle n \rangle \langle \Gamma \rangle} \right) + \text{higher-order terms}
\end{align*}
What EBM picks out as the interaction term should be $\displaystyle -\frac{\langle n \rangle^2}{\langle \Gamma \rangle} \left(2\frac{\delta n \delta \Gamma} {\langle n \rangle \langle \Gamma \rangle} \right)$. Indeed, if we divide the $\fnu$ versus $n_b$ plot into four quadrants (with division lines at $\langle \fnu \rangle$ and $\langle n_b \rangle$), quadrants I and III have negative signs and quadrants II and IV have positive signs, which is consistent with the expression above.

\bibliographystyle{apj}
\bibliography{main}

\begin{thebibliography}{}
\expandafter\ifx\csname natexlab\endcsname\relax\def\natexlab#1{#1}\fi
\providecommand{\url}[1]{\href{#1}{#1}}

\bibitem[{{Adelberger} {et~al.}(2003){Adelberger}, {Steidel}, {Shapley}, \& {Pettini}}]{Adelberger2003}
{Adelberger}, K.~L., {Steidel}, C.~C., {Shapley}, A.~E., \& {Pettini}, M. 2003, \apj, 584, 45

\bibitem[{{Becker} {et~al.}(2019){Becker}, {D'Aloisio}, {Davies}, {Hennawi}, \& {Simcoe}}]{Becker2019}
{Becker}, G., {D'Aloisio}, A., {Davies}, F.~B., {Hennawi}, J.~F., \& {Simcoe}, R.~A. 2019, \baas, 51, 440

\bibitem[{{Becker} \& {Bolton}(2013)}]{Becker2013}
{Becker}, G.~D., \& {Bolton}, J.~S. 2013, \mnras, 436, 1023

\bibitem[{{Becker} {et~al.}(2011){Becker}, {Bolton}, {Haehnelt}, \& {Sargent}}]{Becker2011}
{Becker}, G.~D., {Bolton}, J.~S., {Haehnelt}, M.~G., \& {Sargent}, W. L.~W. 2011, \mnras, 410, 1096

\bibitem[{{Becker} {et~al.}(2015){Becker}, {Bolton}, \& {Lidz}}]{Becker2015}
{Becker}, G.~D., {Bolton}, J.~S., \& {Lidz}, A. 2015, \pasa, 32, e045

\bibitem[{{Bi} \& {Davidsen}(1997)}]{Bi1997}
{Bi}, H., \& {Davidsen}, A.~F. 1997, \apj, 479, 523

\bibitem[{{Boera} {et~al.}(2016){Boera}, {Murphy}, {Becker}, \& {Bolton}}]{Boera2016}
{Boera}, E., {Murphy}, M.~T., {Becker}, G.~D., \& {Bolton}, J.~S. 2016, \mnras, 456, L79

\bibitem[{{Bolton} {et~al.}(2022){Bolton}, {Gaikwad}, {Haehnelt}, {Kim}, {Nasir}, {Puchwein}, {Viel}, \& {Wakker}}]{Bolton2022}
{Bolton}, J.~S., {Gaikwad}, P., {Haehnelt}, M.~G., {et~al.} 2022, \mnras, 513, 864

\bibitem[{Bosman {et~al.}(2018)Bosman, Fan, Jiang, Reed, Matsuoka, Becker, \& Haehnelt}]{Bosman2018}
Bosman, S. E.~I., Fan, X., Jiang, L., {et~al.} 2018, Monthly Notices of the Royal Astronomical Society, 479, 1055.
\newblock \url{https://doi.org/10.1093/mnras/sty1344}

\bibitem[{{Cantalupo} {et~al.}(2012){Cantalupo}, {Lilly}, \& {Haehnelt}}]{Cantalupo2012}
{Cantalupo}, S., {Lilly}, S.~J., \& {Haehnelt}, M.~G. 2012, \mnras, 425, 1992

\bibitem[{{Cen} {et~al.}(1994){Cen}, {Miralda-Escud{\'e}}, {Ostriker}, \& {Rauch}}]{Lya1}
{Cen}, R., {Miralda-Escud{\'e}}, J., {Ostriker}, J.~P., \& {Rauch}, M. 1994, \apjl, 437, L9

\bibitem[{{Christenson} {et~al.}(2023){Christenson}, {Becker}, {D'Aloisio}, {Davies}, {Zhu}, {Boera}, {Nasir}, {Furlanetto}, \& {Malkan}}]{Christenson2023}
{Christenson}, H.~M., {Becker}, G.~D., {D'Aloisio}, A., {et~al.} 2023, \apj, 955, 138

\bibitem[{{Cooray} {et~al.}(2019){Cooray}, {Aguirre}, {Ali-Haimoud}, {Alvarez}, {Appleton}, {Armus}, {Becker}, {Bock}, {Bowler}, {Bowman}, {Bradford}, {Breysse}, {Bromm}, {Burns}, {Caputi}, {Castellano}, {Chang}, {Chary}, {Chiang}, {Cohn}, {Conselice}, {Cuby}, {Davies}, {Dayal}, {Dore}, {Farrah}, {Ferrara}, {Finkelstein}, {Furlanetto}, {Hazelton}, {Heneka}, {Hutter}, {Jacobs}, {Koopmans}, {Kovetz}, {La Piante}, {Le Fevre}, {Liu}, {Ma}, {Ma}, {Malhotra}, {Mao}, {Marrone}, {Masui}, {McQuinn}, {Mirocha}, {Mortlock}, {Murphy}, {Nayyeri}, {Natarajan}, {Nithyanandan}, {Parsons}, {Pello}, {Pope}, {Rhoads}, {Rhodes}, {Riechers}, {Robertson}, {Scarlata}, {Serjeant}, {Saliwanchik}, {Salvaterra}, {Schneider}, {Silva}, {Sahl{\'e}n}, {Santos}, {Switzer}, {Temi}, {Trac}, {Venkatesan}, {Visbal}, {Zaldarriaga}, {Zemcov}, \& {Zheng}}]{Cooray2019}
{Cooray}, A., {Aguirre}, J., {Ali-Haimoud}, Y., {et~al.} 2019, \baas, 51, 48

\bibitem[{{Croft} {et~al.}(1998){Croft}, {Weinberg}, {Katz}, \& {Hernquist}}]{fgpa1}
{Croft}, R. A.~C., {Weinberg}, D.~H., {Katz}, N., \& {Hernquist}, L. 1998, \apj, 495, 44

\bibitem[{{Furlanetto} {et~al.}(2019){Furlanetto}, {Beardsley}, {Carilli}, {Mirocha}, {Aguirre}, {Ali-Haimoud}, {Alvarez}, {Becker}, {Bowman}, {Breysse}, {Bromm}, {Bull}, {Burns}, {Carucci}, {Chang}, {Chiang}, {Cohn}, {Davies}, {DeBoer}, {Dickinson}, {Dillon}, {Dor{\'e}}, {Dvorkin}, {Fialkov}, {Finkelstein}, {Gnedin}, {Hazelton}, {Jacobs}, {Karkare}, {Koopmans}, {Kovetz}, {La Plante}, {Lidz}, {Liu}, {Ma}, {Mao}, {Masui}, {McQuinn}, {Mesinger}, {Munoz}, {Murray}, {Parsons}, {Pober}, {Robertson}, {Sievers}, {Switzer}, {Thyagarajan}, {Trac}, {Visbal}, \& {Zaldarriaga}}]{Furlanetto2019}
{Furlanetto}, S., {Beardsley}, A., {Carilli}, C.~L., {et~al.} 2019, \baas, 51, 142

\bibitem[{{Gaikwad} {et~al.}(2017){Gaikwad}, {Khaire}, {Choudhury}, \& {Srianand}}]{Gaikwad2017}
{Gaikwad}, P., {Khaire}, V., {Choudhury}, T.~R., \& {Srianand}, R. 2017, \mnras, 466, 838

\bibitem[{{Gaikwad} {et~al.}(2021){Gaikwad}, {Srianand}, {Haehnelt}, \& {Choudhury}}]{Gaikwad2021}
{Gaikwad}, P., {Srianand}, R., {Haehnelt}, M.~G., \& {Choudhury}, T.~R. 2021, \mnras, 506, 4389

\bibitem[{{Gaikwad} {et~al.}(2020){Gaikwad}, {Rauch}, {Haehnelt}, {Puchwein}, {Bolton}, {Keating}, {Kulkarni}, {Ir{\v{s}}i{\v{c}}}, {Ba{\~n}ados}, {Becker}, {Boera}, {Zahedy}, {Chen}, {Carswell}, {Chardin}, \& {Rorai}}]{Gaikwad2020}
{Gaikwad}, P., {Rauch}, M., {Haehnelt}, M.~G., {et~al.} 2020, \mnras, 494, 5091

\bibitem[{{Garaldi} {et~al.}(2019){Garaldi}, {Gnedin}, \& {Madau}}]{Garaldi2019}
{Garaldi}, E., {Gnedin}, N.~Y., \& {Madau}, P. 2019, \apj, 876, 31

\bibitem[{{Garaldi} {et~al.}(2022){Garaldi}, {Kannan}, {Smith}, {Springel}, {Pakmor}, {Vogelsberger}, \& {Hernquist}}]{Garaldi2022}
{Garaldi}, E., {Kannan}, R., {Smith}, A., {et~al.} 2022, \mnras, 512, 4909

\bibitem[{{Garaldi} {et~al.}(2023){Garaldi}, {Kannan}, {Smith}, {Borrow}, {Vogelsberger}, {Pakmor}, {Springel}, {Hernquist}, {Gal{\'a}rraga-Espinosa}, {Yeh}, {Shen}, {Xu}, {Neyer}, {Spina}, {Almualla}, \& {Zhao}}]{thesan2}
---. 2023, arXiv e-prints, arXiv:2309.06475

\bibitem[{{Gnedin}(2014)}]{gnedin14}
{Gnedin}, N.~Y. 2014, \apj, 793, 29

\bibitem[{{Gnedin}(2022)}]{Gnedin2022}
---. 2022, \apj, 937, 17

\bibitem[{{Gnedin} \& {Kaurov}(2014)}]{gnedinandkaurov_14}
{Gnedin}, N.~Y., \& {Kaurov}, A.~A. 2014, \apj, 793, 30

\bibitem[{{Gnedin} {et~al.}(2011){Gnedin}, {Kravtsov}, \& {Rudd}}]{gnedin11}
{Gnedin}, N.~Y., {Kravtsov}, A.~V., \& {Rudd}, D.~H. 2011, \apjs, 194, 46

\bibitem[{{Hausen} {et~al.}(2023){Hausen}, {Robertson}, {Zhu}, {Gnedin}, {Madau}, {Schneider}, {Villasenor}, \& {Drakos}}]{Hausen2023}
{Hausen}, R., {Robertson}, B.~E., {Zhu}, H., {et~al.} 2023, \apj, 945, 122

\bibitem[{{Hernquist} {et~al.}(1996){Hernquist}, {Katz}, {Weinberg}, \& {Miralda-Escud{\'e}}}]{Lya2}
{Hernquist}, L., {Katz}, N., {Weinberg}, D.~H., \& {Miralda-Escud{\'e}}, J. 1996, \apjl, 457, L51

\bibitem[{{Hui} \& {Gnedin}(1997)}]{Hui1997}
{Hui}, L., \& {Gnedin}, N.~Y. 1997, \mnras, 292, 27

\bibitem[{{Hui} {et~al.}(1997){Hui}, {Gnedin}, \& {Zhang}}]{fgpa0}
{Hui}, L., {Gnedin}, N.~Y., \& {Zhang}, Y. 1997, \apj, 486, 599

\bibitem[{{Kakiichi} {et~al.}(2018){Kakiichi}, {Ellis}, {Laporte}, {Zitrin}, {Eilers}, {Ryan-Weber}, {Meyer}, {Robertson}, {Stark}, \& {Bosman}}]{Kakiichi2018}
{Kakiichi}, K., {Ellis}, R.~S., {Laporte}, N., {et~al.} 2018, \mnras, 479, 43

\bibitem[{{Kannan} {et~al.}(2022){Kannan}, {Garaldi}, {Smith}, {Pakmor}, {Springel}, {Vogelsberger}, \& {Hernquist}}]{thesan}
{Kannan}, R., {Garaldi}, E., {Smith}, A., {et~al.} 2022, \mnras, 511, 4005

\bibitem[{{Kashino} {et~al.}(2023){Kashino}, {Lilly}, {Matthee}, {Eilers}, {Mackenzie}, {Bordoloi}, \& {Simcoe}}]{Kashino2023}
{Kashino}, D., {Lilly}, S.~J., {Matthee}, J., {et~al.} 2023, \apj, 950, 66

\bibitem[{{Keating} {et~al.}(2020){Keating}, {Weinberger}, {Kulkarni}, {Haehnelt}, {Chardin}, \& {Aubert}}]{Keating2020}
{Keating}, L.~C., {Weinberger}, L.~H., {Kulkarni}, G., {et~al.} 2020, \mnras, 491, 1736

\bibitem[{{La Plante} {et~al.}(2019){La Plante}, {Alvarez}, {Fialkov}, {Aguirre}, {Ali-Ha{\"\i}moud}, {Becker}, {Bowman}, {Breysse}, {Bromm}, {Bull}, {Burns}, {Cappelluti}, {Carucci}, {Chang}, {Cleary}, {Cooray}, {Chen}, {Chiang}, {Cohn}, {DeBoer}, {Dillon}, {Dor{\'e}}, {Dvorkin}, {Ferraro}, {Furlanetto}, {Hazelton}, {Hill}, {Jacobs}, {Karkare}, {Keating}, {Koopmans}, {Kovetz}, {Lidz}, {Liu}, {Ma}, {Mao}, {Masui}, {McQuinn}, {Mirocha}, {Mu{\~n}oz}, {Murray}, {Parsons}, {Pober}, {Saliwanchik}, {Sievers}, {Thyagarajan}, {Trac}, {Vikhlinin}, {Visbal}, \& {Zaldarriaga}}]{LaPlante2019}
{La Plante}, P., {Alvarez}, M., {Fialkov}, A., {et~al.} 2019, \baas, 51, 394

\bibitem[{Lou {et~al.}(2013)Lou, Caruana, Gehrke, \& Hooker}]{ebm}
Lou, Y., Caruana, R., Gehrke, J., \& Hooker, G. 2013, in Proceedings of the 19th ACM SIGKDD international conference on Knowledge discovery and data mining, 623--631

\bibitem[{{Meyer} {et~al.}(2019){Meyer}, {Bosman}, {Kakiichi}, \& {Ellis}}]{Meyer2019}
{Meyer}, R.~A., {Bosman}, S. E.~I., {Kakiichi}, K., \& {Ellis}, R.~S. 2019, \mnras, 483, 19

\bibitem[{{Meyer} {et~al.}(2020){Meyer}, {Kakiichi}, {Bosman}, {Ellis}, {Laporte}, {Robertson}, {Ryan-Weber}, {Mawatari}, \& {Zitrin}}]{Meyer2020}
{Meyer}, R.~A., {Kakiichi}, K., {Bosman}, S. E.~I., {et~al.} 2020, \mnras, 494, 1560

\bibitem[{{Miralda-Escud{\'e}} {et~al.}(2000){Miralda-Escud{\'e}}, {Haehnelt}, \& {Rees}}]{Miralda-Escude2000}
{Miralda-Escud{\'e}}, J., {Haehnelt}, M., \& {Rees}, M.~J. 2000, \apj, 530, 1

\bibitem[{{Mo} \& {White}(1996)}]{Mo1996}
{Mo}, H.~J., \& {White}, S.~D.~M. 1996, \mnras, 282, 347

\bibitem[{{Nasir} \& {D'Aloisio}(2020)}]{Nasir2020}
{Nasir}, F., \& {D'Aloisio}, A. 2020, \mnras, 494, 3080

\bibitem[{{Nori} {et~al.}(2019){Nori}, {Jenkins}, {Koch}, \& {Caruana}}]{nori2019}
{Nori}, H., {Jenkins}, S., {Koch}, P., \& {Caruana}, R. 2019, arXiv e-prints, arXiv:1909.09223

\bibitem[{{Ocvirk} {et~al.}(2016){Ocvirk}, {Gillet}, {Shapiro}, {Aubert}, {Iliev}, {Teyssier}, {Yepes}, {Choi}, {Sullivan}, {Knebe}, {Gottl{\"o}ber}, {D'Aloisio}, {Park}, {Hoffman}, \& {Stranex}}]{coda1}
{Ocvirk}, P., {Gillet}, N., {Shapiro}, P.~R., {et~al.} 2016, \mnras, 463, 1462

\bibitem[{{Ocvirk} {et~al.}(2020){Ocvirk}, {Aubert}, {Sorce}, {Shapiro}, {Deparis}, {Dawoodbhoy}, {Lewis}, {Teyssier}, {Yepes}, {Gottl{\"o}ber}, {Ahn}, {Iliev}, \& {Hoffman}}]{coda2}
{Ocvirk}, P., {Aubert}, D., {Sorce}, J.~G., {et~al.} 2020, \mnras, 496, 4087

\bibitem[{{Peeples} {et~al.}(2010){Peeples}, {Weinberg}, {Dav{\'e}}, {Fardal}, \& {Katz}}]{Peeples2010}
{Peeples}, M.~S., {Weinberg}, D.~H., {Dav{\'e}}, R., {Fardal}, M.~A., \& {Katz}, N. 2010, \mnras, 404, 1281

\bibitem[{{Rakic} {et~al.}(2012){Rakic}, {Schaye}, {Steidel}, \& {Rudie}}]{Rakic2012}
{Rakic}, O., {Schaye}, J., {Steidel}, C.~C., \& {Rudie}, G.~C. 2012, \apj, 751, 94

\bibitem[{{Rauch}(1998)}]{Rauch1998}
{Rauch}, M. 1998, \araa, 36, 267

\bibitem[{{Rieke} {et~al.}(2019){Rieke}, {Arribas}, {Bunker}, {Charlot}, {Finkelstein}, {Maiolino}, {Robertson}, {Willott}, {Windhorst}, {Eisenstein}, {Nelson}, {Tacchella}, {Egami}, {Endsley}, {Frye}, {Hainline}, {Hviding}, {Rieke}, {Williams}, {Willmer}, \& {Woodrum}}]{Rieke2019}
{Rieke}, M., {Arribas}, S., {Bunker}, A., {et~al.} 2019, \baas, 51, 45

\bibitem[{{Telikova} {et~al.}(2019){Telikova}, {Shternin}, \& {Balashev}}]{Telikova2019}
{Telikova}, K.~N., {Shternin}, P.~S., \& {Balashev}, S.~A. 2019, \apj, 887, 205

\bibitem[{{Walther} {et~al.}(2019){Walther}, {O{\~n}orbe}, {Hennawi}, \& {Luki{\'c}}}]{Walther2019}
{Walther}, M., {O{\~n}orbe}, J., {Hennawi}, J.~F., \& {Luki{\'c}}, Z. 2019, \apj, 872, 13

\bibitem[{{Weinberg} {et~al.}(2003){Weinberg}, {Dav{\'e}}, {Katz}, \& {Kollmeier}}]{fgpa2}
{Weinberg}, D.~H., {Dav{\'e}}, R., {Katz}, N., \& {Kollmeier}, J.~A. 2003, in American Institute of Physics Conference Series, Vol. 666, The Emergence of Cosmic Structure, ed. S.~H. {Holt} \& C.~S. {Reynolds}, 157--169

\bibitem[{{Zhang} {et~al.}(1997){Zhang}, {Anninos}, {Norman}, \& {Meiksin}}]{Lya3}
{Zhang}, Y., {Anninos}, P., {Norman}, M.~L., \& {Meiksin}, A. 1997, \apj, 485, 496

\end{thebibliography}

\end{CJK*}
\end{document}